\documentclass[twocolumn]{aastex631}
\usepackage{graphicx} 

\begin{document}
\title{3D MC I: X-ray Tomography Begins to Unravel the 3-D Structure of a Molecular Cloud in our Galaxy's Center}

\submitjournal{\apjl}

\correspondingauthor{Cara Battersby}
\email{cara.battersby@uconn.edu}

\author[0000-0001-6776-2550]{Samantha W. Brunker}
\affiliation{University of Connecticut, Department of Physics, 196A Auditorium Road, Unit 3046, Storrs, CT 06269, USA}

\author[0000-0002-6073-9320]{Cara Battersby}
\affiliation{University of Connecticut, Department of Physics, 196A Auditorium Road, Unit 3046, Storrs, CT 06269, USA}

\author[0009-0005-9578-2192]{Danya Alboslani}
\affiliation{University of Connecticut, Department of Physics, 196A Auditorium Road, Unit 3046, Storrs, CT 06269, USA}

\author[0000-0003-0724-2742]{Ma\"ica Clavel}
\affiliation{Univ. Grenoble Alpes, CNRS, IPAG, F-38000 Grenoble, France}

\author[0000-0001-7330-8856]{Daniel L. Walker}
\affiliation{University of Connecticut, Department of Physics, 196A Auditorium Road, Unit 3046, Storrs, CT 06269, USA}
\affiliation{ALMA Regional Centre Node, Jodrell Bank Centre for Astrophysics, The University of Manchester, Manchester M13 9PL, UK}

\author[0000-0002-5776-9473]{Dani Lipman}
\affiliation{University of Connecticut, Department of Physics, 196A Auditorium Road, Unit 3046, Storrs, CT 06269, USA}

\author[0000-0003-0946-4365]{H Perry Hatchfield }
\affiliation{University of Connecticut, Department of Physics, 196A Auditorium Road, Unit 3046, Storrs, CT 06269, USA}
\affiliation{Jet Propulsion Laboratory, California Institute of Technology, 4800 Oak Grove Drive, Pasadena, CA, 91109, USA}

\author[0000-0002-8219-4667]{R\'egis Terrier }
\affiliation{APC, Universit\'e Paris Diderot, CNRS/IN2P3, CEA/Irfu, Observatoire de Paris, Sorbonne Paris Cit\'e, 10 rue Alice Domon et L\'eonie Duquet, 75205 Paris Cedex 13, France}

\begin{abstract}
Astronomers have used observations of the Galactic gas and dust via infrared, microwave, and radio to study molecular clouds in extreme environments such as the Galactic center.  More recently, X-ray telescopes have opened up a new wavelength range in which to study these molecular clouds. Previous flaring events from SgrA* propagate X-rays outwards in all directions, and these X-rays interact with the surrounding molecular gas, illuminating different parts of the clouds over time.  We use a combination of X-ray observations from Chandra and molecular gas tracers (line data from Herschel and the Submillimeter Array) to analyze specific features in the Sticks cloud, one of three clouds in the Three Little Pigs system in the Central Molecular Zone (Galactic longitude and latitude of 0.106 and -0.082 degrees respectively).  We also present a novel X-ray tomography method we used to create  3-D map of the Sticks cloud. By combining X-ray and molecular tracer observations, we are able to learn more about the environment inside the Sticks cloud. 
\end{abstract}

\section{Introduction}

Stars are born deep within clouds of molecular gas in interstellar space \citep{Heyer2015}.  Since the discovery of MCs in 1970 \citep{Wilson1970}, astronomers have been seeking to understand and map the distribution of molecular clouds across the Galaxy and beyond \citep{RomanDuval2010,Sun2020}. MCs are detected via molecular gas as well as dust absorption and emission.  Starting with early CO maps \citep{Wilson1970,Perault1985}, to the most recent high-resolution maps using a wide variety of other molecular tracers \citep{Edenhofer2024,Zucker2022,Alves2020,Zucker2020}, the structure and environments of molecular clouds in the Milky Way has been of particular interest because the process of star formation is tightly linked with the 3-D density structure of these MCs \citep{Hacar2023}.  Despite the abundance of data for MCs in the Galaxy, creating 3-D models of individual clouds is challenging.  Pioneering new work has made tremendous progress creating 3-D models of nearby MCs by mapping out dust extinction \citep{Edenhofer2024,Zucker2021,Zucker2020,Zucker2018,Zucker2019}, however, understanding the 3-D structure of MCs outside our solar neighborhood has remained elusive.

This work focuses on a single cloud in the Central Molecular Zone (CMZ), the Sticks cloud, which was named as such in the CMZoom survey \citep{Battersby2020}.  The Sticks cloud is one of three clouds in a system named the Three Little Pigs located in the CMZ with a Galactic longitude and latitude of 0.106 and -0.082 degrees respectively. The Sticks cloud is an average Galactic Center molecular cloud \citep{Battersby2020,Callanan2023}, with properties of a median dust temperature of 22 Kelvin, a radius of 1.7 parsecs, a mass of 2.2$\times 10^{4}$ solar masses, and a median column density of 1$\times 10^{23}$ cm$^{-2}$.  All are well within the normal range of Galactic Center molecular clouds \citep{Battersby2025}.  In this work, we present a novel method for investigating the 3-D structure of MCs in the Galactic Center using X-ray tomography, and we also compare the X-ray observations with molecular tracers to study the structure and properties of the cloud in more detail.  

\begin{figure*}[ht]
\centering
\includegraphics[width=\linewidth]{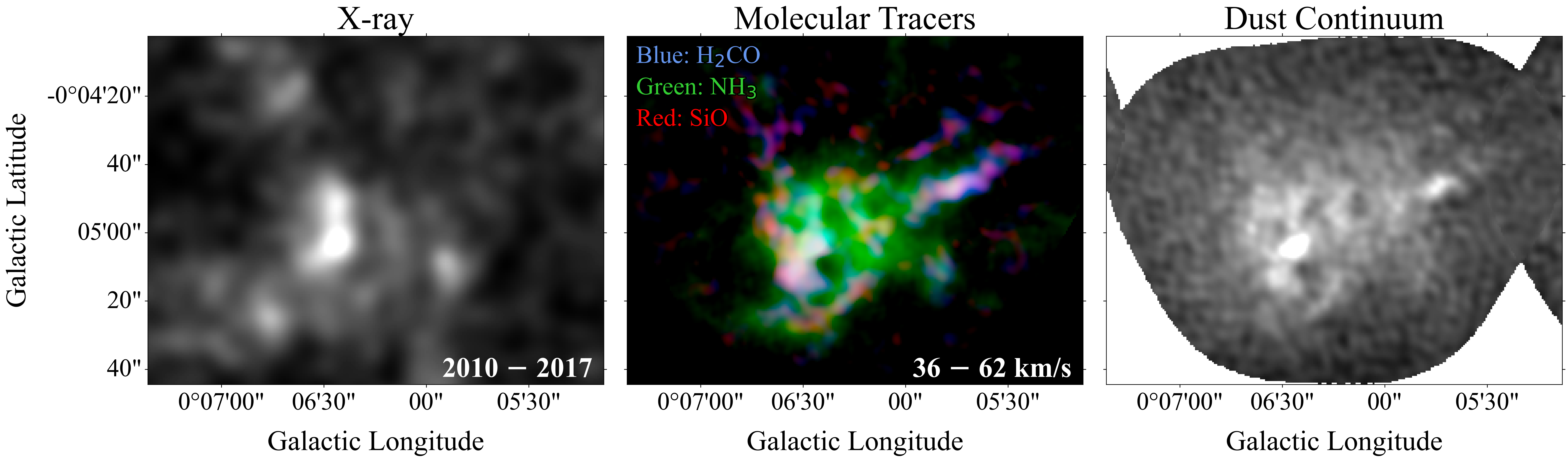}
\caption{A comparison of the multi-wavelength observations of the Sticks cloud used in this analysis shows good correlation between the X-ray echoes and the position of the molecular gas and dust. \textit{Left}: Integrated X-ray observations of 6.4 keV emission from 2010 to 2017 with Gaussian smoothing applied (see methods for details). \textit{Center}: RGB image incorporating observations of H$_{2}$CO (blue), NH$_{3}$ (green), and SiO (red). Each molecular tracer image used to create the RGB image was integrated over the velocity range of the Sticks cloud, 36$-$62 km s$^{-1}$. \textit{Right}: CMZoom 1.3mm dust continuum observation of the Sticks cloud. An interactive version of these panels is provided and shows the individual X-ray observations for each year as well as the individual molecular components that make up the RGB image.} 
\label{fig:multiwavelength_comp}
\end{figure*}

\section{Data}

\subsection{Chandra X-ray Observations}

When the supermassive black hole at the center of our Galaxy has an intense accretion event, it produces X-rays which travel outwards and interact with the molecular gas in the CMZ. Some of the X-rays are absorbed by neutral iron atoms in the molecular clouds, and the iron atoms then reemit at 6.4 keV (referred to as an X-ray echo), which can be observed with modern X-ray telescopes such as Chandra and XMM-Newton \citep{Clavel2013,Ponti2010}.  As the X-rays travel through the molecular cloud over time, they illuminate the material, and we observe individual slices of the cloud through discrete observations. Using observations taken over a period of several years, we use these X-ray cross-sectional slices to generate a 3-D model.  The time axis of the X-ray data can be converted to position using the geometry of the black hole-cloud system as well as information about the X-ray burst.  Chandra X-ray observations of the Sticks cloud exist for almost every year between 1999 – 2017 (excluding 2012 and 2014) and allow us to detect the X-ray echo corresponding to when X-rays from SgrA* interacted with the molecular gas in the cloud \citep{Clavel2013,Chuard2018}.  We see the resulting fluorescence in the Sticks cloud starting in the observations from 2010 and the propagation of the light through the cloud for all the remaining observations through 2017.

X-ray observations of the CMZ performed in the last three decades have revealed a strongly variable X-ray emission, correlated with molecular clouds, that has been interpreted as X-ray echoes from past activity from the supermassive black hole, Sgr A*. The X-ray signal associated with this reflection process is characterized by a power-law continuum component created by Compton scattering that is absorbed at low energy.  On top of the continuum there are emission lines, including the 6.4 keV one created by the fluorescence of neutral iron, which dominates the spectrum. For a given incident luminosity and cloud distance, the flux measured in the 6.4 keV line is directly proportional to the density of the illuminated material \citep{Capelli2012,Tsuru2014} and can therefore be used to trace the matter distribution within each cloud.

\begin{figure*}[ht]
\centering
\includegraphics[width=\linewidth]{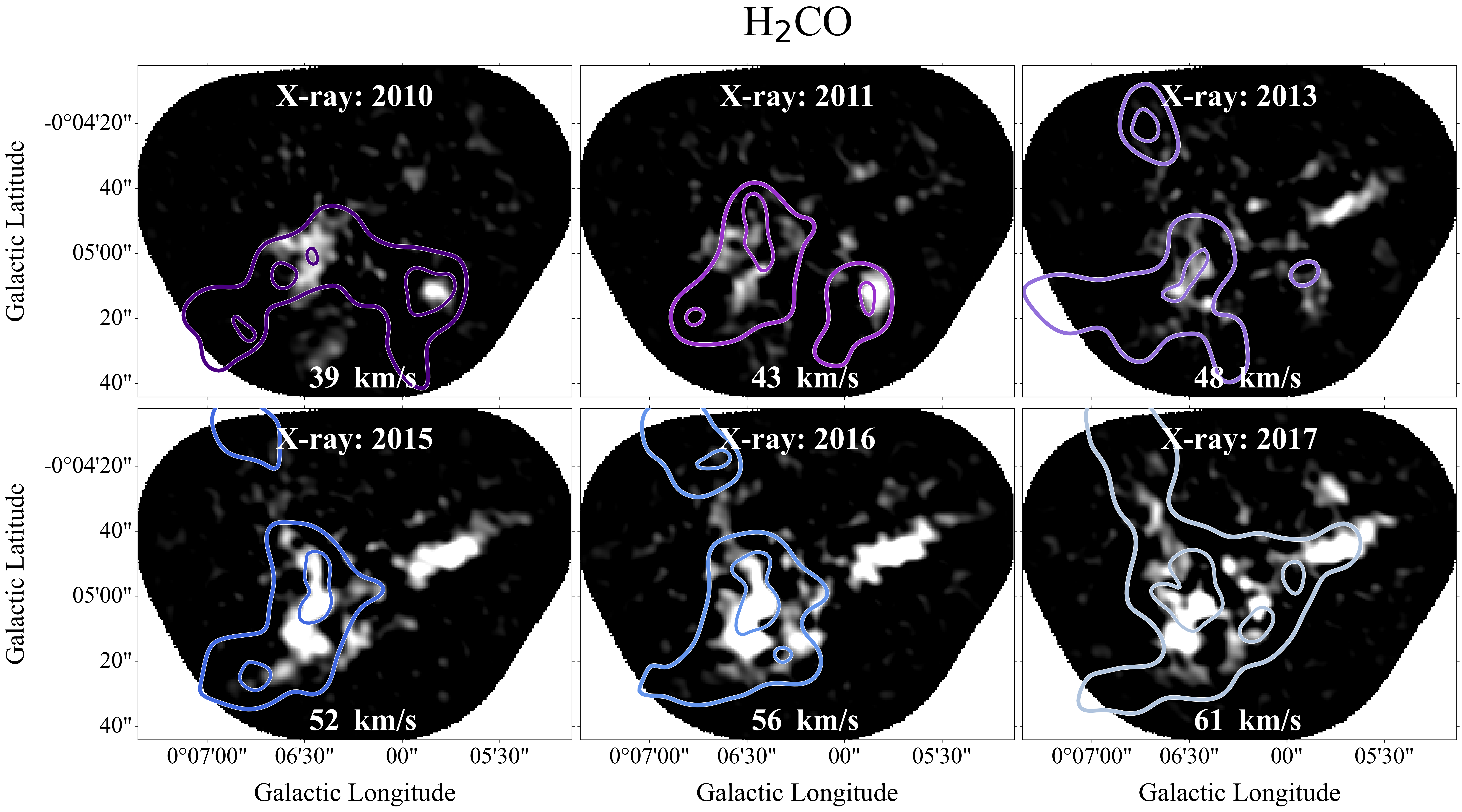}
\caption{A comparison between each year of X-ray observations (colored contours) and consecutive H$_{2}$CO integrated velocity slices (grayscale background) suggests a good match in the overall morphological structures. Each panel shown has the average velocity indicated at the bottom of the image. The contours for each year of X-ray observations were matched with an integrated H$_{2}$CO image sequentially.  By eye, we see relatively good agreement between the X-ray and molecular gas features especially in the bottom three panels.  Unsurprisingly, the sequential comparison between X-ray echoes and velocity slices isn't perfect, though we do see some strikingly good correlation for the small-scale structure in years 2013, 2015, and 2016 especially.} 
\label{fig:xray_h2co_comp}
\end{figure*}

In this work we focus on the X-ray echoes towards the Sticks cloud, also called Br2 or Bridge cloud in previous X-ray work \citep{Clavel2013,Chuard2018,Marin2023,Ponti2010,Sunyaev1998,Churazov2017}. We use all Chandra ACIS-I data covering the Sticks cloud obtained between 2010 and 2017 (Chandra observation IDs : 11843, 12949, 13438, 13508, 13016, 13017, 14897, 14941, 14942, 17236, 17239, 17237, 17240, 18852, 17238, 17241, 20118, 20807, 20808). The data were reduced using ciao v.4.8, and we produced the continuum subtracted 6.4 keV maps following the method described in \citet{Clavel2013}. We merged the individual observations based on the observing date in order to obtain one map per year. The exposure of the data set is not uniform across the period considered. There are no observations in 2012 and 2014. There are rather shallow exposures in 2010 ($\sim$80ks centered on Sgr A*, i.e. off-axis and therefore less sensitive in our region of interest) and in 2013 (50ks centered on the molecular clouds plus $\sim$40ks centered on Sgr A*). There is a deeper coverage in 2011, 2015, 2016 and 2017 (with at least $\sim$150ks centered on the clouds). These yearly maps were then smoothed using a Gaussian smoothing kernel, to mask noise fluctuations in all X-ray maps. We evaluated smoothing kernels ranging from two to ten, and ultimately, we chose a smoothing kernel of 4 as it was the smallest kernel that gave a sufficient signal-to-noise and removed spurious noise, while still retaining the small-scale structure of the cloud. If the smoothing kernel used is too large, the X-ray features are smoothed out to the point where it is impossible to compare to the molecular data. Contours corresponding to X-ray data smoothed with a Gaussian smoothing kernel of 4 are shown later in Figures~\ref{fig:xray_h2co_comp}, \ref{fig:3Dmodel}, and \ref{fig:colden}. To look at the higher resolution X-ray peaks inside the main body of the cloud, we used X-ray data smoothed using a Gaussian smoothing kernel of 3. Contours corresponding to this smaller smoothing kernel are shown later in Figures~\ref{fig:xray_h2co_comp} and \ref{fig:3Dmodel}.

\subsection{Molecular Data}

The X-ray observations are discrete, and in order to generate a 3-D model using X-ray tomography, we first need to confirm that the X-ray features are indeed echoes, and therefore, correspond to features seen in the molecular cloud.  The molecular line tracers we chose to compare with the X-ray data are H$_{2}$CO (the 3(0,3)$-$2(0,2) transition at a rest frequency of 218.2 GHz) from the CMZoom survey \citep{Battersby2020,Callanan2023} (uses observations from the Submillimeter Array) which traces dense gas, SiO (5$-$4 transition at a rest frequency of 217.1 GHz) also from the CMZoom survey \citep{Battersby2020,Callanan2023} which traces shocked gas, and NH$_{3}$ (the (1,1) species with a rest wavelength of 23.69 GHz) from VLA observations \citep{Butterfield2022} which also traces dense gas.  For the molecular data sets, the peak emission of the Sticks cloud is seen in the velocity range of 36-62 km s$^{-1}$.  In addition to the molecular line data, we also used other data sets such as the Herschel column density observations \citep{Barnes2017,Battersby2011} and the dust continuum data from the CMZoom survey \citep{Battersby2020} (mapped at 230 GHz) which correlates to the gas column density.  The three panels in Figure~\ref{fig:multiwavelength_comp} show the Sticks cloud via X-ray echo (6.4 keV emission) observations from 2010$-$2017 (left panel) integrated, three molecular tracers integrated over the velocity range of the cloud (middle panel), and dust continuum (right panel).  The location of the peak flux in the three panels and the general shape of the emission seems to match well between the three panels.

\section{Analysis}
\subsection{Positional comparison between X-ray and molecular datasets}
The visual match between the integrated X-ray and molecular gas emission is promising, however, to verify that the X-ray emission is the result of interactions with the molecular cloud, we perform a quantitative comparison between the images. For each image comparison, we calculate the overlap percentage (OP) between the detected emission in two data sets (DS1 and DS2) defined as 
\begin{equation}
\text{OP}[\text{DS1,DS2}]_{\text{primary}} = \frac{(\text{DS1} \cap \text{DS2})}{\text{total pixel value}},
\end{equation}
where DS1$\cap$DS2 is the intersection between DS1 and DS2, or the number of pixels where emission is detected in both images. The total pixel value is the number of pixels with detected emission in the primary data set which is chosen to be either DS1 or DS2. We use the X-ray as our primary data set as it is the simplest baseline for comparison: while the molecular gas or dust continuum-detected cloud might be more extensive, the X-ray emission is restricted to the portion of the cloud to which all datasets are sensitive. You can also think of the overlap percentage where X-ray is the primary data set as the percent of pixels inside the X-ray contours that overlap with or are covered by the secondary data set. We find overlap percentages of OP[X-ray,H$_{2}$CO]$_{\text{X-ray}}$ = 48\% and OP[X-ray,dust]$_{\text{X-ray}}$ = 54\% for the molecular gas (H$_{2}$CO) and dust continuum respectively. We integrated the X-ray data over the years where we see fluorescence in the cloud (2010-2017), and we integrated the H$_{2}$CO data over the velocity range of the cloud (36-62 km s$^{-1}$).  We use X-ray contours at a level of 3$\times$10$^{-9}$ counts s$^{-1}$ cm$^{-2}$ pixel$^{-1}$, for the H$_{2}$CO data we use contours at a 3-sigma level, and for the dust continuum we use contour levels set to 5-sigma.  If we instead consider the alternate dataset the primary, we find OP[X-ray,H$_{2}$CO]$_{\text{H$_{2}$CO}}$ = 76\% and OP[X-ray,dust]$_{\text{dust}}$ = 84\%. These high OPs verify the high likelihood that the X-ray emission is tracing the same material detected in the molecular gas and dust continuum.  

\begin{figure*}
\centering
\begin{minipage}[h]{1\linewidth}
\centering
\includegraphics[width=0.55\linewidth]{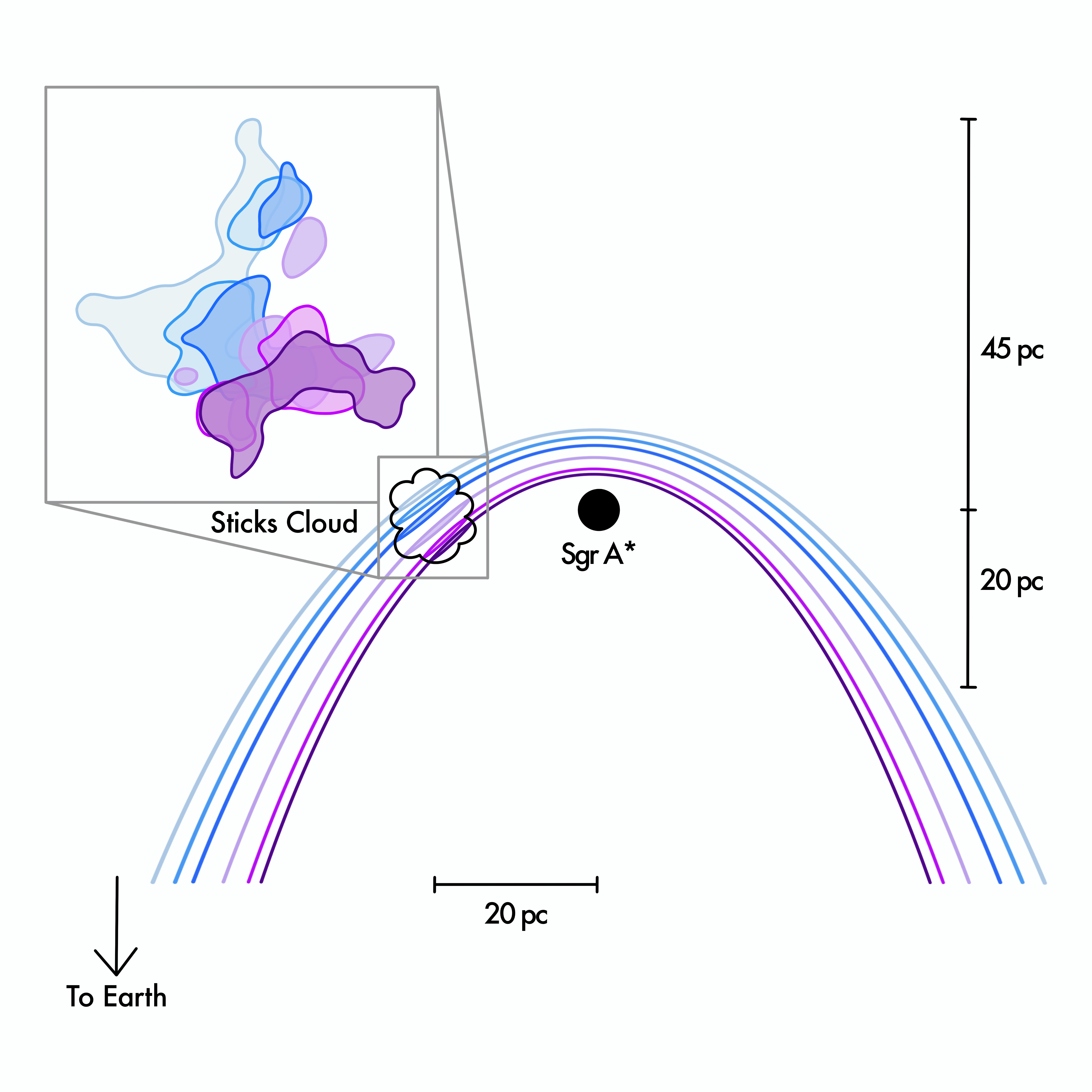} 
\end{minipage}
\vfill
\vspace{0.01 cm}
\begin{minipage}[h]{0.47\textwidth}
\centering
\includegraphics[width=0.68\linewidth]{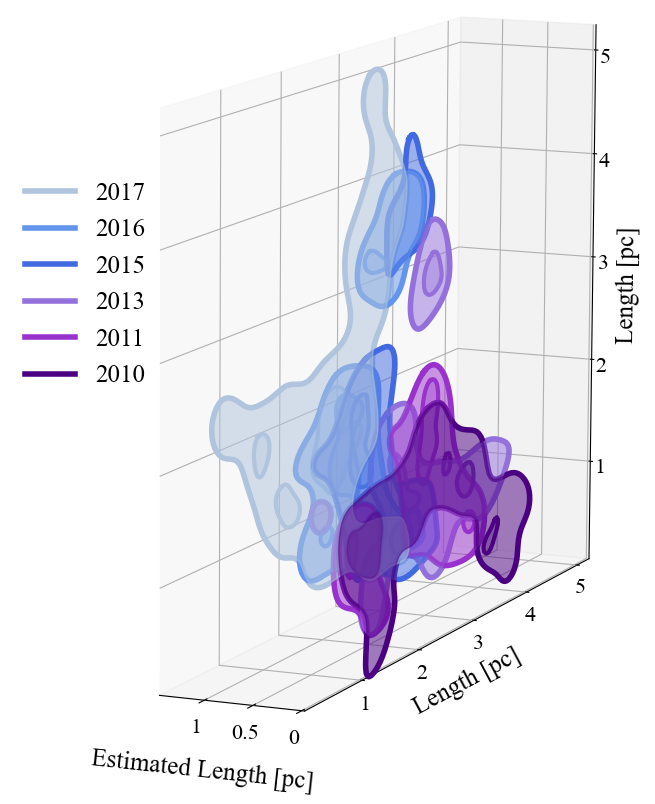}
\end{minipage}
\hspace{0 cm}
\begin{minipage}{0.47\textwidth}
\centering
\includegraphics[width=0.85\linewidth]{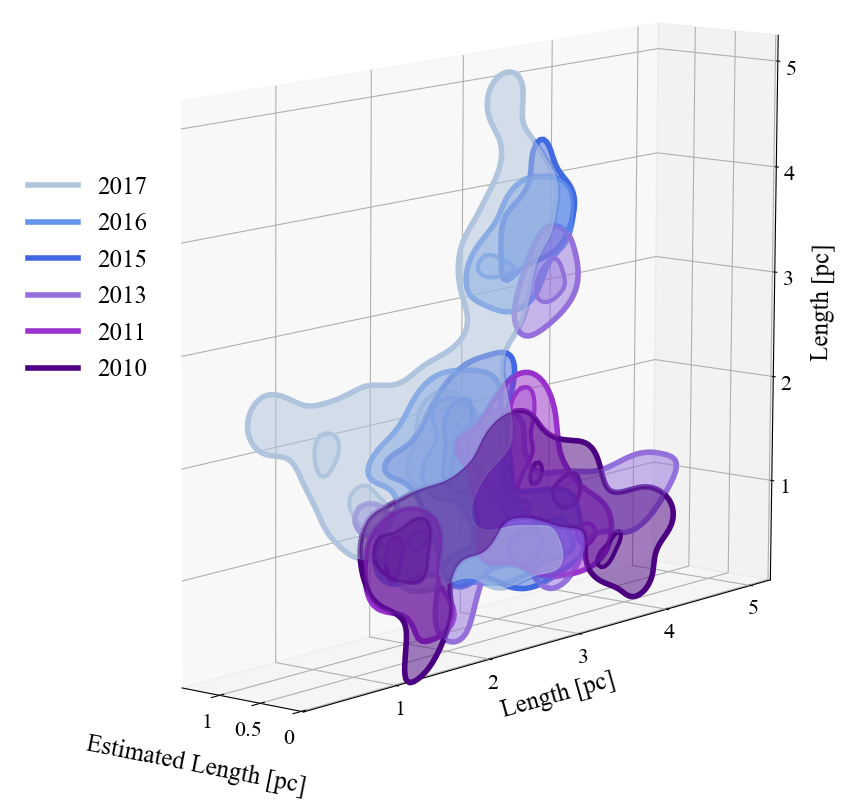}
\end{minipage}
\caption{[Top] An illustration (not to scale) of the top-down view of the geometry of the Sticks cloud and SgrA*.  The separation between the Sticks cloud and the black hole is around 20 pc, but the position of the cloud either in front of or behind the black hole is less certain.  The scale bar on the right side of the illustration indicates the range of distances the Sticks cloud could be at in front of or behind SgrA*.  The colored parabolas represent the X-rays traveling outwards from the black hole, reflecting off the cloud, and traveling back to our observatories at Earth, and we observe the X-rays at discrete times.  We use these slices in our tomography method to create a 3-D model of the cloud, which is shown in the zoom-in panel.  An animated version of this illustration is available. [Bottom] 3-D model of the Sticks cloud using the X-ray tomography method shown at two different rotation angles. The contour colors and widths are the same as shown in Figure~\ref{fig:xray_h2co_comp} The axes have all been converted to parsecs to give an estimation of the size scale of the Sticks cloud.  The physical distance between the X-ray years was calculated assuming the Sticks cloud is 25 parsecs behind the central black hole, along the line of sight, with an illuminating event age of 200 years \citep{Marin2023}.  This gives an approximate length of 1.25 parsecs between years 2010 and 2017.  Conservatively, the Sticks cloud could be in the range of 20 parsecs in front of the black hole to 45 parsecs behind the black hole based on X-ray observations \citep{Clavel2013}, which would give a range of lengths from 1.18$-$1.63 parsecs for the total z-axis. The line-of-sight of the Sticks cloud is shown from right to left (from the 2010 contours to the 2017 contours).  A rotating version of this figure is also provided.}
\label{fig:3Dmodel}
\end{figure*}

We can take the comparison of the integrated X-ray and molecular gas emission in the Sticks cloud a step further.  Instead of looking at the integrated data, we can compare the individual years of X-ray observations with the progression of the molecular line data in velocity space. We start with the simplest assumption that the gas velocities in the cloud trace its structure in 3-D and therefore, that sequential velocity slices would line up with sequential years of X-ray observations. This assumption may be valid in very simplistic geometries, where for example the cloud is simply moving along an orbital stream, and its velocity increases monotonically with increasing distance, but there are many factors which complicate the velocity structure of a MC (i.e., star formation, turbulence).  The comparison of the X-ray with the H$_{2}$CO data is shown in Figure~\ref{fig:xray_h2co_comp}.  We divided the H$_{2}$CO observations equally within the velocity range of the cloud to make each of the integrated images shown in Figure~\ref{fig:xray_h2co_comp}, and we assigned the contours for each of the X-ray observations sequentially to the H$_{2}$CO images.  The two contour levels shown for the X-ray data correspond to two different sets of Gaussian smoothing levels and flux limits, to highlight both the dense cores and the more diffuse emission surrounding them.  The thicker exterior contours for each year correspond to the data with a Gaussian smoothing kernel of 4 pixels and a contour level of 3$\times$10$^{-9}$ counts s$^{-1}$ cm$^{-2}$ pixel$^{-1}$.  The thinner interior contours for each year correspond to the data with a Gaussian smoothing kernel of 3 pixels and a contour level of 7$\times$10$^{-9}$ counts s$^{-1}$ cm$^{-2}$ pixel$^{-1}$.  

We assessed through visual inspection how well the locations of peak flux and general shape of the X-ray contours matched with the H$_{2}$CO data, and we computed an overlap percentage for each comparison for both contour levels.  These overlap percentages are shown in the Figures~\ref{fig:overlap_kernel4} and \ref{fig:overlap_kernel3} in the Appendix, where the blue text starting with “X:” is the overlap percentage with X-ray as the primary data set, and the pink text starting with “H:” is the overlap percentage with H$_{2}$CO as the primary data set.  The text colors correspond to the colors the X-ray and H$_{2}$CO contours are plotted with.  We find that the overlap percentages for the X-ray contours associated with the dense cores are, on average, 27\% higher than the X-ray contours associated with the diffuse gas.  These high OPs for the densest cores indicate that the X-ray and molecular observations are likely probing the same material.

In addition to comparing the X-ray contours to the H$_{2}$CO molecular data, shown in Figure~\ref{fig:xray_h2co_comp}, we also did this comparison with NH$_{3}$ and SiO, shown in Appendix Figures~\ref{fig:xray_nh3_comp} and \ref{fig:xray_sio_comp}.  Similar to the H$_{2}$CO data, we divided the NH$_{3}$ and SiO observations equally within the velocity range of the cloud such that each integrated image consists of four averaged velocity slices shown in Figures~\ref{fig:xray_nh3_comp} and \ref{fig:xray_sio_comp} with the average velocity of the four observations shown at the bottom of each image. The assigned X-ray years for each integrated image match those used for the H$_{2}$CO analysis.  The X-ray contours for each year are plotted in the same colors as shown in Figures~\ref{fig:xray_h2co_comp} and \ref{fig:3Dmodel}with the same smoothing kernels and contour levels used.   

The uncertainties associated with equating structures seen in position-position-velocity (PPV) space to position-position-position (PPP) space \citep{Beaumont2013} is an important reason why we primarily use the X-ray data to build our 3-D model of the Sticks cloud.  Because the velocity structure of a cloud can be complicated and we don't expect the PPV emission to strictly map to PPP, we don't expect all the X-ray contours to match the molecular velocity data sequentially.  However, we do see the same features present in the molecular data in several of the X-ray year slices. Some areas of spectacularly good agreement are the central cores in 2013, 2015, and 2016 as well as the diffuse northern tail in 2017. The overall morphological agreement, and in particular, the association of the densest regions in both X-ray and molecular line data is striking and is the first time it has been shown on such a small scale.  Now that we have established that the X-ray fluorescence is associated with the structures seen in molecular gas, we can use the X-ray data to construct a 3-D model of the cloud.

\subsection{Using X-ray tomography to create a 3-D model of an MC}

We use the X-ray tomography method which consists of using cross-sectional slices to produce a 3-D model of the Sticks cloud.  The X-ray tomography method is effective in this case thanks to the extensive X-ray coverage over time and because the singular X-ray eruption at the origin of this signal was very short, lasting no more than two years \citep{Clavel2013,Chuard2018,Churazov2017}.  At each given time, the light front from this past event is therefore illuminating a very thin slice of material following a paraboloid that moves away from Sgr A* with time \citep{Sunyaev1998}.Therefore, the geometry of the echo allows us to reconstruct the 3-D distribution of the material within the cloud \citep{Tsuru2014}. The illustration shown in the top half of Figure~\ref{fig:3Dmodel} shows the location of illuminated material (along the parabola) as the X-rays travel outwards from SgrA* and interact with the Sticks cloud.  The colored parabolas represent the location of the X-ray fluorescence for each year that we observed the Sticks cloud (2010-2017) from dark purple to light blue.  The line-of-sight distance of the illuminated material for each observation from 2010-2017 can be determined using the paraboloid equation, 

\begin{equation}
z(t) = \frac{1}{2}\left(ct - \frac{{d_{proj}}^{2}}{ct}\right)
\end{equation}

where t is the age of the illuminating event, c is the speed of light and d$_{\text{proj}}$ is the distance between SgrA* and the cloud in projection \citep{Sunyaev1998}.  The position of the Sticks cloud is not known exactly, so the possible range of distances from the black hole are indicated with the scale bar on the right side of the illustration.  Based on the X-ray observations, we don’t see any shift in the X-ray fluorescence from side-to-side in the cloud, so we know that the X-rays are primarily from front-to-back or back-to-front through the cloud.    
Using the X-ray tomography method, we can create a 3-D model in PPP space as the RA and Dec coordinates can be converted to physical units using basic geometry and the known distance to the GC, while the X-ray year can be converted to distance since we know it moves through the cloud at the speed of light.  This method is illustrated in the zoom-in panel of the illustration in Figure~\ref{fig:3Dmodel}, and the resulting model of the Sticks cloud is shown in the two lower panels of the figure.  A rotating version of this plot is also available.  

For the model shown in Figure~\ref{fig:3Dmodel}, we assume the distance of the cloud from SgrA* is 25 parsecs \citep{Marin2023}, and the illuminating event has an age of 200 years \citep{Marin2023} based on recent X-ray polarization measurements. The X-ray contours for each year are plotted in the same colors as shown in Figure~\ref{fig:xray_h2co_comp} with the same Gaussian smoothing kernels and contour levels used.  This 3-D model likely does not show the entirety of the Sticks cloud, as indicated by the lack of a tapered shape at either end of our model, but due to the correlation between the X-ray echoes and the locations of the densest parts of the cloud observed via molecular tracers, we are able to construct a 3-D model of an interior portion of the Sticks cloud.

\begin{figure*}[ht]
\centering
\includegraphics[width=\linewidth]{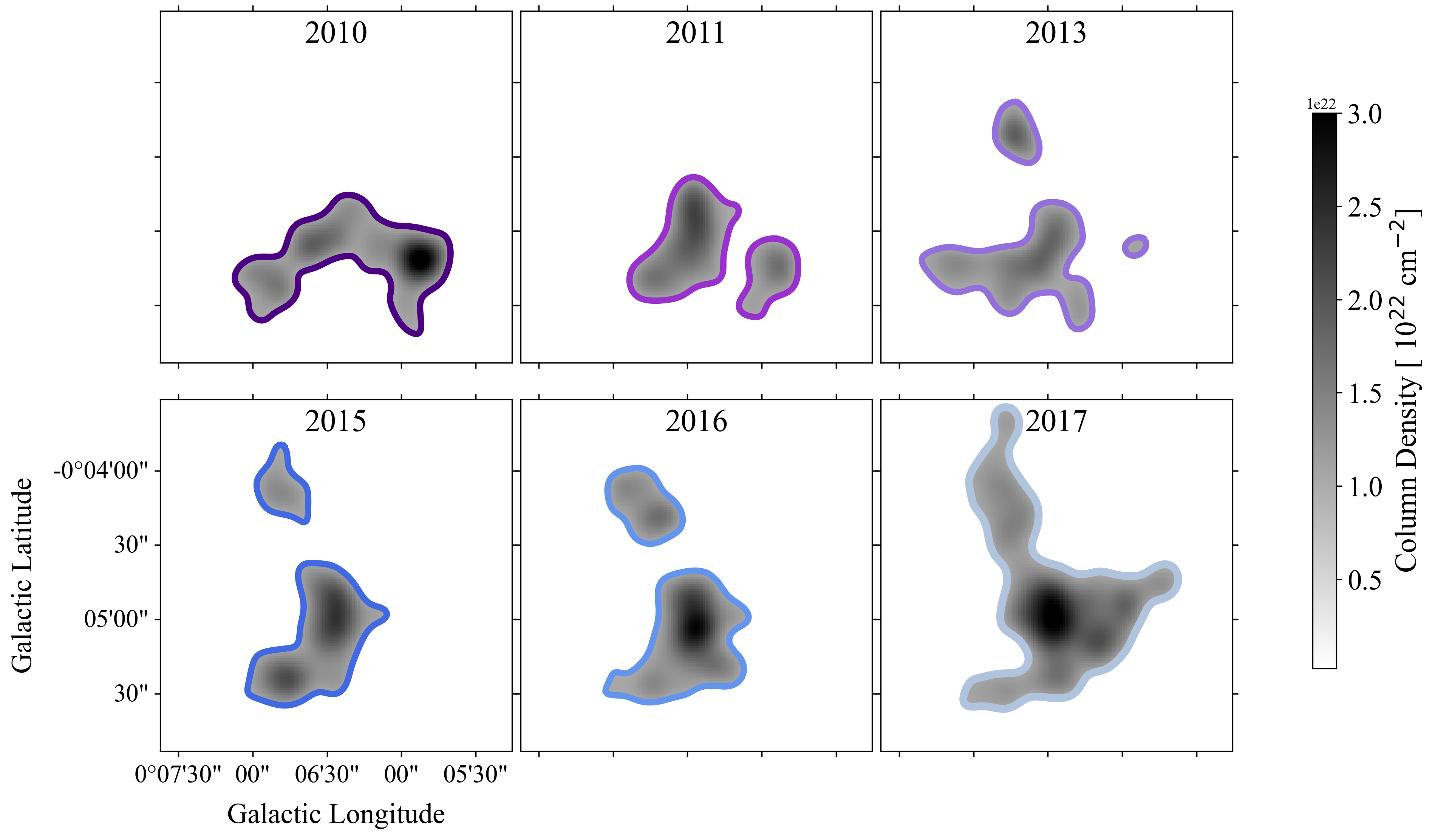}
\caption{The column density calculated for each year of X-ray observations of the Sticks cloud.  The contours shown for each year are the same ones plotted in the previous figures.  We see a variety of dense cores throughout the cloud in the X-ray observations.} 
\label{fig:colden}
\end{figure*}

In this work, we assume that all the X-ray signal is associated with the Sticks cloud, and that, on the cloud scale, the paraboloid can be approximated by a plane that moves with the speed of light. This speed is correct if the Sticks cloud is in the plane of Sgr A*, but it could be faster (slower) if the cloud is in front of (behind) Sgr A*. Current estimation of the age of the event illuminating this cloud range between about 100 and 200 years old, and a new X-ray polarization measurement gives the position of the Sticks cloud as approximately 25 pc behind the black hole \citep{Marin2023}.  The physical space between each year shown in Figure~\ref{fig:3Dmodel} was calculated using these values and gives a physical separation between years of 0.18 parsecs with a total width of the 2010$-$2017 X-ray slices of 1.25 parsecs.  Conservatively, the Sticks cloud could be in the range of 20 parsecs in front of the black hole to 45 parsecs behind the black hole \citep{Clavel2013}, which means the width of the z-axis shown in Figure~\ref{fig:3Dmodel} could range from 1.18$-$1.63 parsecs.

\subsection{Density distribution of the Sticks cloud}

In addition to using the X-ray observations to create a 3-D model of the Sticks cloud, we can use a combination of X-ray and Herschel data to create a 3-D density map.  The density distribution of a MC is a critical input for many theoretical models \citep{Klessen2011,Federrath2016}.  We assume that the X-ray flux in each slice is proportional to the column density in this slice \citep{Capelli2012,Tsuru2014} which allows us to calculate a density normalization factor for the Sticks cloud.  We started by creating an integrated X-ray image using the relevant years and obtaining Herschel column density maps of the cloud \citep{Barnes2017,Battersby2011}. For both the X-ray and Herschel data, we found the position of peak column density and peak X-ray flux within the Sticks cloud and a normalization factor was calculated by dividing the peak column density by the peak X-ray flux, since we are assuming they are proportional.  We then created column density maps for each X-ray slice by multiplying each pixel by the column density normalization, shown in Figure~\ref{fig:colden}.  The resulting maps serve as an upper limit on the column density, since without infinitely continuous coverage it is possible to miss material and overestimate the normalization value using this method.  The locations and sizes of the densest regions matches well with the molecular observations of dense gas tracers.  This method offers another use for the X-ray observations of MCs, and the resulting density distributions can be used in further analyses.

\section{Conclusions}
In this work we have shown that the X-ray echoes observed in the Sticks cloud are correlated to the molecular gas and dust in the cloud, and that we are likely observing X-rays from SgrA* interacting with the densest regions of a MC in the Galactic center.   Using X-ray tomography (combining cross-sectional slices), we have produced a 3-D model of a portion of the Sticks cloud.  We also used a combination of X-ray and Herschel observations to calculate the column densities of the gas associated with each slice of our 3-D model.  Combining the 3-D model and density map of the Sticks cloud with known physical properties of the MC will help us to better understand the environment inside the cloud and how it affects star formation processes. We were fortuitous that Chandra observed the X-ray echoes of the Sticks MC, enabling 3-D tomography of this MC in our extreme Galactic center. 

\appendix
\restartappendixnumbering

\section{Extended data figures}
\begin{figure*}[ht]
\centering
\includegraphics[width=\linewidth]{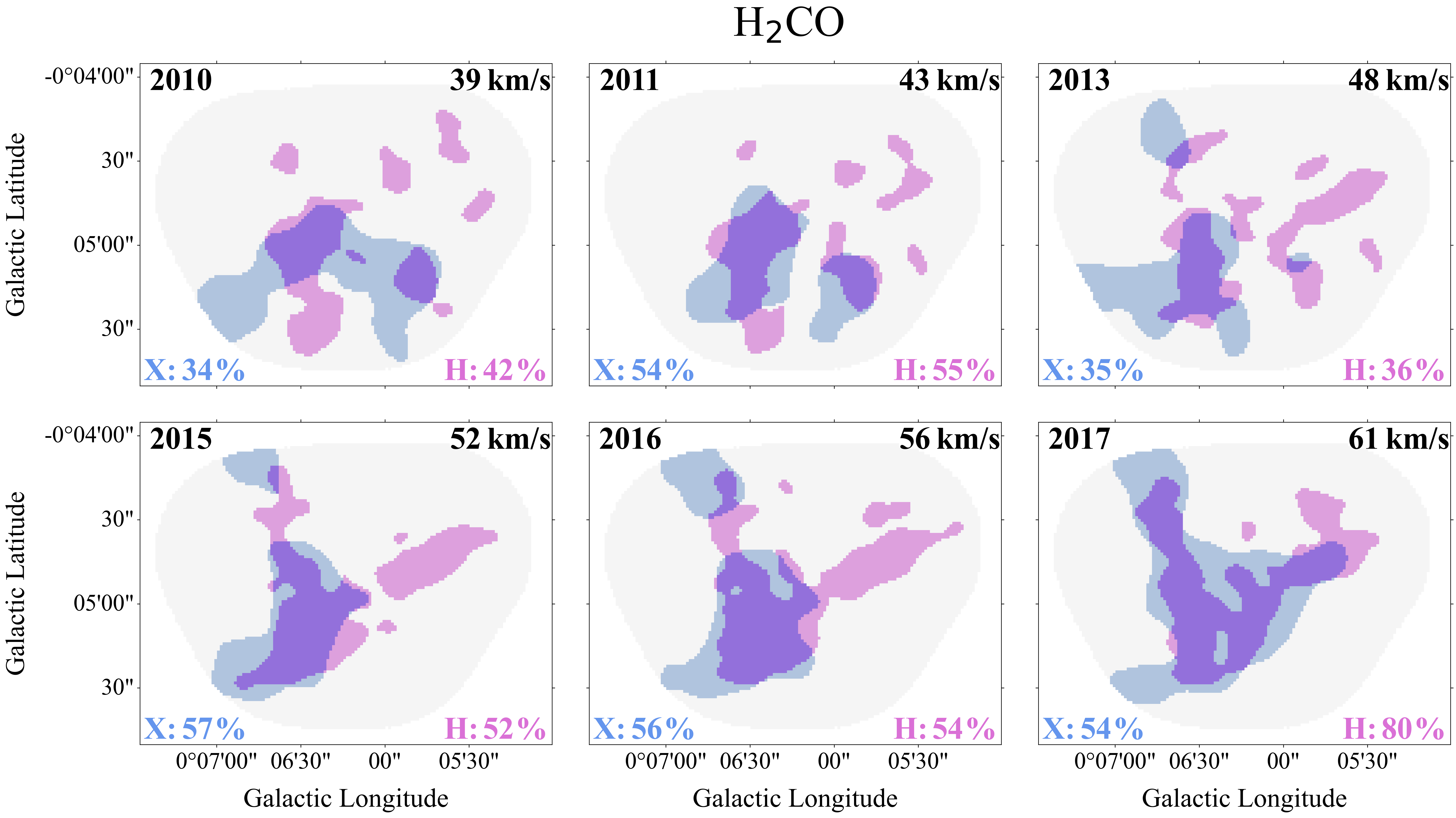}
\caption{A visual representation of the overlap percentage calculation done for the individual years of X-ray observations (given in the upper left of each panel) and the H$_{2}$CO observations divided into the same velocity ranges shown in Figure~\ref{fig:xray_h2co_comp} (given in the upper right of each panel).   The blue-filled shapes are the X-ray contours at a level of 3$\times$10$^{-9}$ counts s$^{-1}$ cm$^{-2}$ pixel$^{-1}$ shown as the outer contours in previous figures, and they trace even the diffuse X-ray emission from the cloud. The pink-filled shapes are the 3-sigma H$_{2}$CO contours.  The purple regions show where the two sets of contours overlap.  The overlap percentages with X-ray being the primary data set are shown in blue in the lower left of each panel, and the percentages with H$_{2}$CO being the primary data set are shown in pink in the lower right of each panel.} 
\label{fig:overlap_kernel4}
\end{figure*}

\begin{figure*}[ht]
\centering
\includegraphics[width=\linewidth]{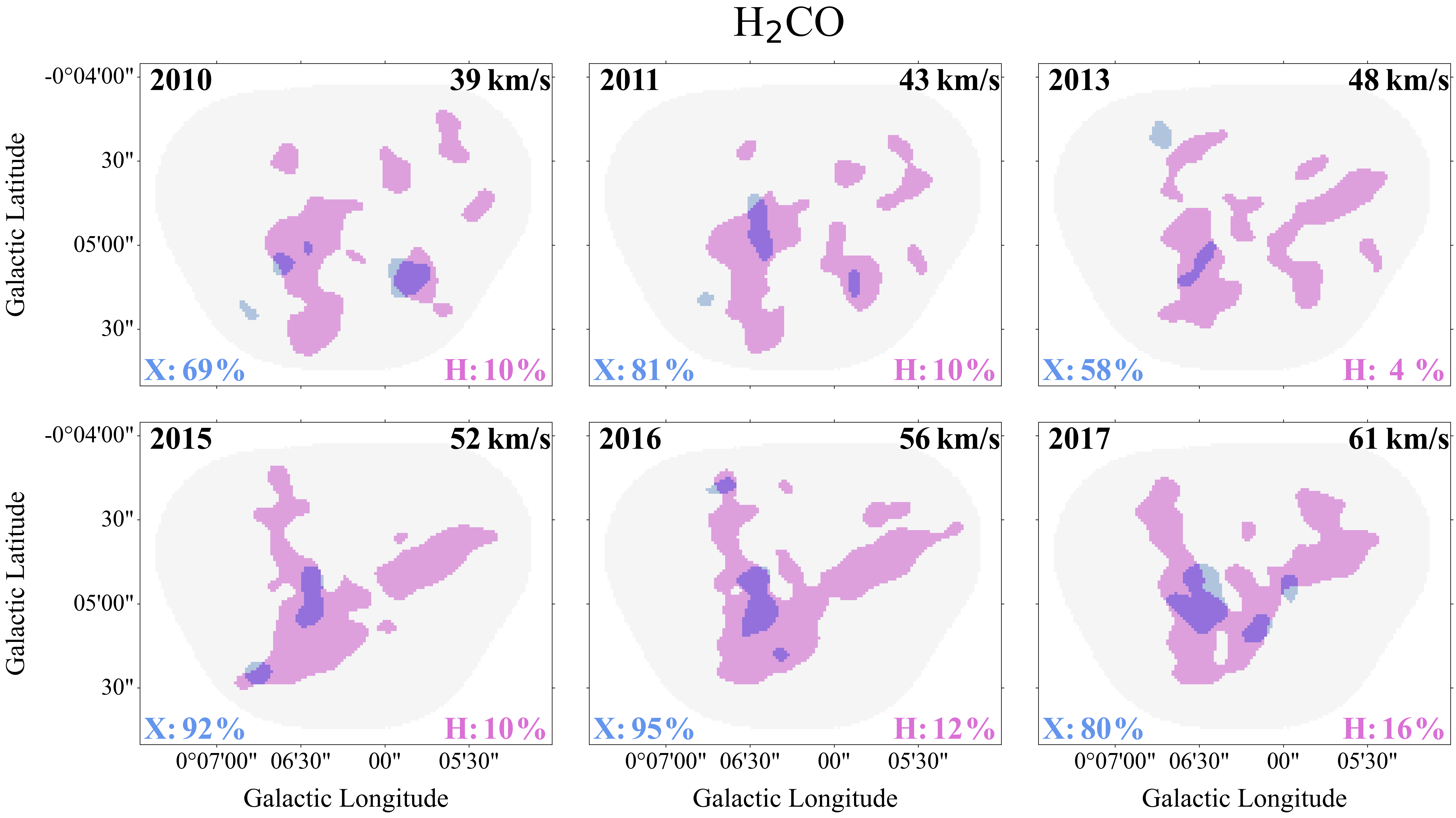}
\caption{This is the same type of comparison as shown in Extended Data Figure 3, however the blue-filled shapes are the X-ray contours at a level of 7$\times$10$^{-9}$ counts s$^{-1}$ cm$^{-2}$ pixel$^{-1}$ also shown in previous figures as the interior contours, and they trace the densest parts of the cloud.  The overlap percentages with X-ray being the primary data set are shown in blue in the lower left of each panel, and the percentages with H$_{2}$CO being the primary data set are shown in pink in the lower right of each panel. These percentages for the dense cores of the cloud are, on average, 27\% higher than the percentages shown in Figure~\ref{fig:overlap_kernel4}, which makes sense as H$_{2}$CO is a dense gas tracer.} 
\label{fig:overlap_kernel3}
\end{figure*}

\begin{figure*}[ht]
\centering
\includegraphics[width=\linewidth]{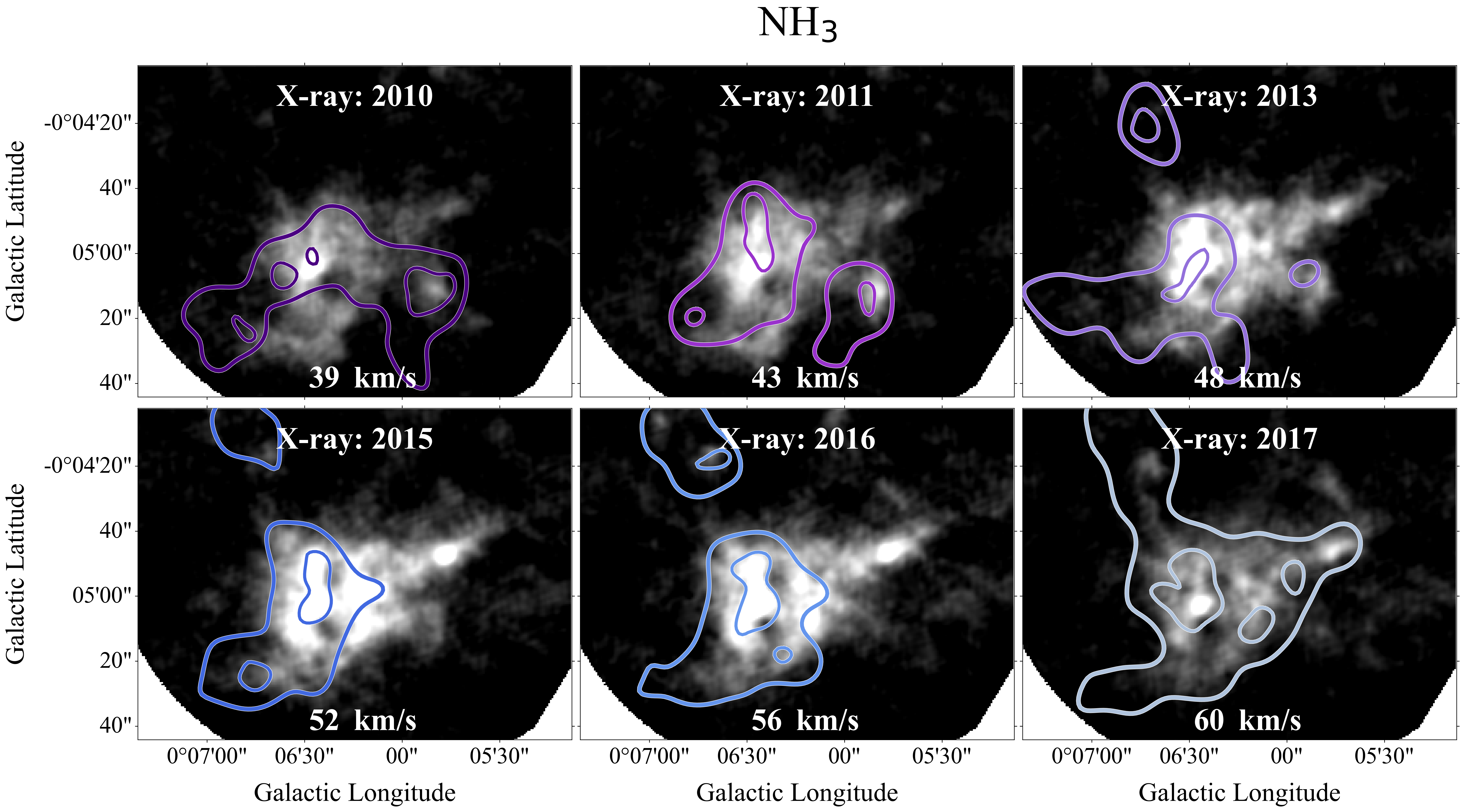}
\caption{A comparison between each year of X-ray data and NH$_{3}$ integrated images made using an equal number of consecutive velocity slices contained within the full velocity range of the Sticks cloud. Each panel shown has the average velocity indicated at the bottom of the image. The contours for each X-ray year were matched with an integrated NH$_{3}$ image sequentially. We see relatively good agreement between the X-ray and molecular gas features especially in the bottom three panels. The contour colors and widths are the same as shown in Figure~\ref{fig:xray_h2co_comp} This analysis is the same as that shown in Figure~\ref{fig:xray_h2co_comp} using NH$_{3}$ instead of H$_{2}$CO.} 
\label{fig:xray_nh3_comp}
\end{figure*}

\begin{figure*}[ht]
\centering
\includegraphics[width=\linewidth]{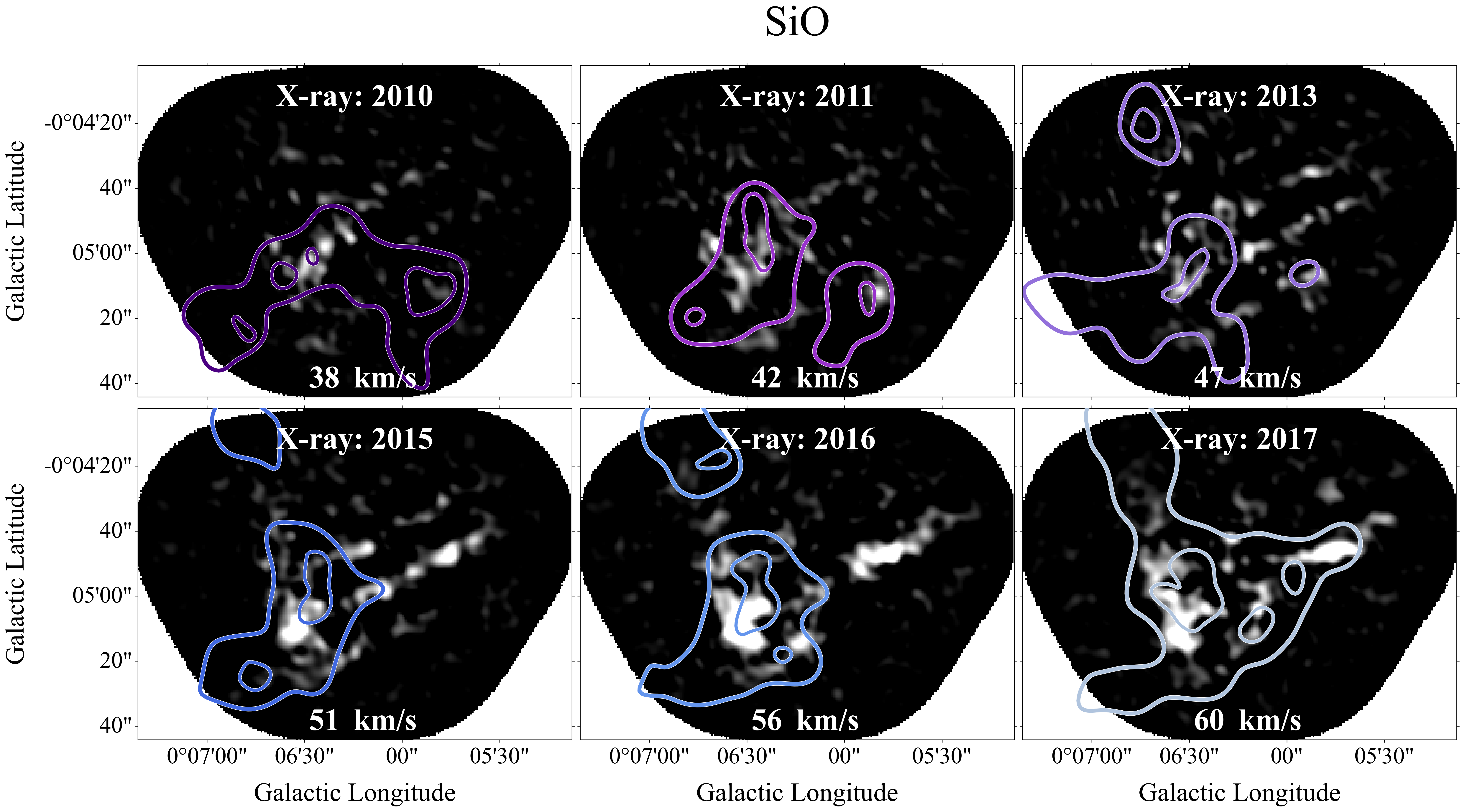}
\caption{A comparison between each year of X-ray data and SiO integrated images made using an equal number of consecutive velocity slices contained within the full velocity range of the Sticks cloud. Each panel shown has the average velocity indicated at the bottom of the image. The contours for each X-ray year were matched with an integrated SiO image sequentially. We see relatively good agreement between the X-ray and molecular gas features especially in the bottom three panels. The contour colors and widths are the same as shown in Figure~\ref{fig:xray_h2co_comp} This analysis is the same as that shown in Figure~\ref{fig:xray_h2co_comp} and \ref{fig:xray_nh3_comp} but using SiO.} 
\label{fig:xray_sio_comp}
\end{figure*}

\begin{acknowledgments}
SWB, CB, and DA gratefully acknowledge funding from the National Aeronautics and Space Administration through the Astrophysics Data Analysis Program under Award “3-D MC: Mapping Circumnuclear Molecular Clouds from X-ray to Radio,” Grant No. 80NSSC22K1125. Additionally, CB gratefully acknowledges funding from National Science Foundation under Award Nos. 2108938, 2206510, and CAREER 2145689. DA acknowledges the Summer Undergraduate Research Fund Award and Research Travel Award from the Office of Undergraduate Research at the University of Connecticut as well as the FAMOUS Travel Grant from the American Astronomical Society. MC acknowledges financial support from the Centre National d’Etudes Spatiales (CNES).

The scientific results reported in this article are based to a significant degree on observations made by the Chandra X-ray Observatory. This research also made use of data from the Submillimeter Array and the authors wish to recognize and acknowledge the very significant cultural role and reverence that the summit of Maunakea has always had within the indigenous Hawaiian community and we are most fortunate to have had the opportunity to utilize observations from this mountain.
\end{acknowledgments}

\software{This research has made use of software provided by the Chandra X-ray Center (CXC) in the application package CIAO. This work also made use of Astropy3: a community-developed core Python package and an ecosystem of tools and resources for astronomy \citep{Astropy2013, Astropy2018, Astropy2022}, SAO ds9 \citep{Joye2003}, 736 and Matplotlib \citep{Hunter2007}.}

\end{document}